\documentclass[journal=jctcce,manuscript=article,reqno]{achemso}

\usepackage{siunitx,amsmath,amssymb,graphicx,rotating,algorithm,algorithmicx,algpseudocode,setspace}
\usepackage[version=3]{mhchem} 
\usepackage{epstopdf}
\usepackage{multirow}
\usepackage{longtable}
\usepackage{color}
\usepackage[table]{xcolor}
\usepackage{booktabs}
\usepackage{textcomp}
\usepackage{subcaption}
\usepackage{bm}
\usepackage{mathrsfs} 
\usepackage{physics} 
\usepackage{breqn} 
\usepackage{svg}
\usepackage[export]{adjustbox}
\usepackage{amsmath}
\usepackage{siunitx}
\usepackage{xcolor}
\usepackage{chemfig}
\usepackage{caption}

\graphicspath{{figures/}} 

\author{Thomas Knoll}
\affiliation{Department of Chemistry and Institute for Advanced Computational Science, Stony Brook University, Stony Brook, NY 11794}

\author{Benjamin G. Levine}
\affiliation{Department of Chemistry and Institute for Advanced Computational Science, Stony Brook University, Stony Brook, NY 11794}
\email{ben.levine@stonybrook.edu}
\phone{631-632-2381}

\title[An \textsf{achemso} demo]
  {Simulating Electron Dynamics with GPU-Accelerated Real-Time Tamm-Dancoff Approximation}


\begin{document}





\begin{tocentry}
\includegraphics[width=1.0\linewidth]{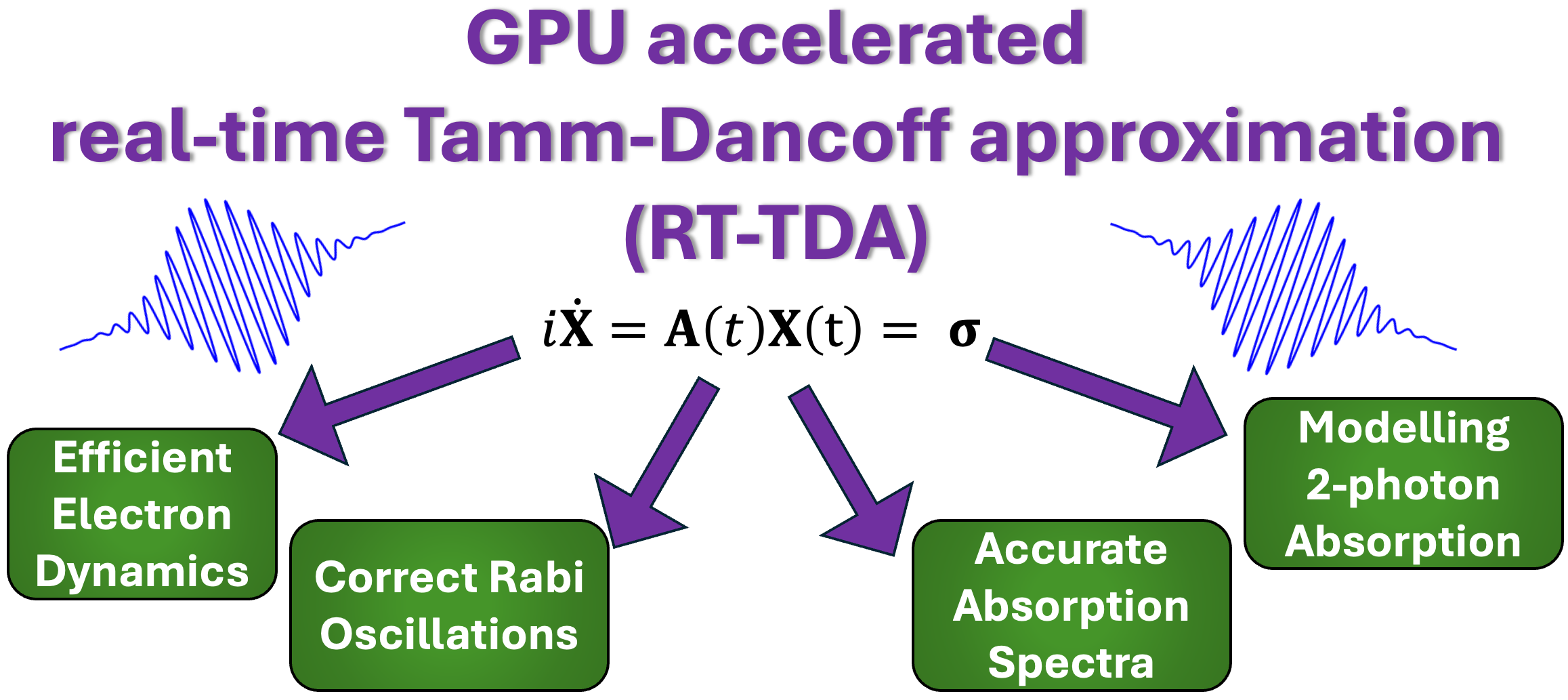}
\end{tocentry}

\begin{abstract}
Time-dependent electronic structure methods provide an efficient, accurate, and robust alternative to traditional time dependent methods for computing both linear and non-linear optical properties.
With this in mind, we have developed the real-time Tamm-Dancoff approximation (RT-TDA). 
This is an approach to model electron dynamics by propagating the linear-response time-dependent density functional theory (LR-TDDFT) amplitudes within the Tamm-Dancoff approximation (TDA) and adiabatic approximation. 
Because the electronic structure is propagated in real-time in a many-electron basis, RT-TDA overcomes known limitations of adiabatic Kohn-Sham RT-TDDFT for describing dynamics in intense fields. 
Acceleration by graphic processing units (GPUs) enables simulations of larger molecules and on longer timescales.
To demonstrate the utility of our approach, we present the calculations of the linear absorption spectrum of a large organic molecule (120 heavy atoms), of Rabi oscillations, and of nonlinear 2-photon absorption, in which we observe the AC Stark effect.
\end{abstract}


\section{Introduction}
\label{sec:intro}

Understanding electron dynamics is central to modern photochemistry and materials science, as it governs energy transfer, nonlinear optics, and laser control of chemical reactivity.\cite{Zewail2000, Krausz2009, Nisoli2017} 
However, simulating these ultrafast processes remains a significant challenge due to the need for methods that are both computationally efficient and accurate. 
In recent years, time-dependent electronic structure methods\cite{Kosloff1988, Li2005, Chapman2011, Nascimento2016, Nascimento2017, Goings2017, Li2020} have emerged as powerful tools for modeling such phenomena, providing access to excitation spectra\cite{Nascimento2016, Nascimento2017}, population dynamics\cite{Padmanaban2008, Wang2013, Mosquera2015, GrobasIllobre2021, Durden2022}, and nonlinear responses\cite{Takimoto2007, Chapman2011, Carlstroem2022}.
One reason for their popularity is that nonlinear effects may be computed by explicitly incorporating the laser field into the working equation.\cite{Luppi2012, NguyenDang2014, Schoenborn2016, Peng2018}
In addition, time-dependent methods can be very efficient for computing linear spectra of large systems, especially when a large number of electronically excited states are of interest.\cite{Tussupbayev2015, Nascimento2016, Nascimento2017}
Furthermore, time-dependent frameworks bypass the need to diagonalize the Hamiltonian, and instead rely on direct time propagation. Replacing potentially unstable diagonalization with a relatively robust initial value problem enables high-throughput calculations of excited-state absorptions spectra, for example.\cite{Mehmood2024}

Real-time time-dependent density functional theory\cite{Runge1984, gross1996_2, marques_2006, Maitra2016} (RT-TDDFT) is a widely used time-dependent electronic structure method due to its desirable balance between accuracy and cost.
Time-dependent configuration interaction (TDCI), with or without an active space expansion, is another popular approach.\cite{Cederbaum1999, Krause2005, Rohringer2006, Krause2007, Schlegel2007, Greenman2010, Ulusoy2011, Hochstuhl2012, Luppi2012, Bauch2014, NguyenDang2014, White2015,  Wang2017, Peng2018, Liu2019, Carlstroem2022, Schlegel2023, Witzorky2024, Durden2025}
Both methods are accurate and effective in many cases, but both also have limitations.
In RT-TDDFT, an incorrect description of Rabi oscillations\cite{fuks_2011, Habenicht2014} and unphysical peak shifting\cite{Provorse2015} have been observed.
Both of these artifacts arise from a dynamic detuning present due to the nonlinear nature of the Fock-like Kohn-Sham operator.
However, RT-TDDFT captures the system's response to external fields to all orders, extending beyond the limitations of the perturbative regime.\cite{Bertsch2000, Li2005, Lopata2011, Andrade2012, Luppi2012}
In principle, RT-TDDFT is an exact theory, thus knowing the exact exchange-correlation functional would solve the above mentioned problems.\cite{Runge1984}
In practice, however, the vast majority of TDDFT calculations (whether in the linear response or real time regimes) use the adiabatic approximation, in which the history dependence of the electron density is neglected.
In linear response (LR-)TDDFT, this results in a frequency independent exchange-correlation kernel, and due to this, only single excitations are accessible to LR-TDDFT within the adiabatic approximation.
Despite this, LR-TDDFT\cite{gross1996_1, Casida1995} is a successful method for computing the electronic structure of excited states in a time-independent framework.
It has been extensively benchmarked and consistently demonstrates strong agreement with both experimental data\cite{Leang2012} and high-level wave-function based methods, such as coupled cluster,\cite{Shao2019, Sarkar2021, Jacquemin2009} relative to its computational cost. The benchmarking data includes excitation energies and oscillator strengths.

The Tamm-Dancoff approximation (TDA) is a commonly used simplification of LR-TDDFT in which the de-excitation terms are neglected\cite{Hirata1999}, resulting in a Hermitian eigenvalue problem that is both computationally more efficient and numerically more stable.
A drawback of the TDA is that it violates the Thomas–Reiche–Kuhn (f-sum) rule, which might affect the accuracy of oscillator strengths.\cite{Rocca2014}
However, the TDA has been shown to provide reliable excitation energies, particularly for singlet states and large systems.\cite{Chantzis2013}
In the context of X-ray and core-level spectroscopies, the TDA has also demonstrated accurate and robust performance.\cite{Fransson2024}
In some cases, such as when triplet instabilities are present, it was shown that the TDA provides more accurate results than solving the full Casida equation.\cite{Rangel2017}
A recent extension to the TDA even allows accurate description of conical intersections between the ground and first excited electronic states,\cite{xu_2025} a known failure of LR-TDDFT.\cite{levine_2006}
These findings support the use of TDA as a practical approximation for many applications.

In TDCI, the electronic coefficients \textbf{C} are propagated in a many-electron basis, according to the time-dependent Schrödinger equation, given by
\begin{equation}
i\dot{\textbf{C}}(t) = \textbf{H}(t)\textbf{C}(t),
\label{eq:tdse_coefficients}
\end{equation}
where \textbf{H} is the Hamiltonian matrix.
The propagation in the many-electron basis avoids the issues that RT-TDDFT has with the dynamic detuning.
Here, the nonlinear Fock-operator (or matrix) is replaced with the proper linear Hamiltonian operator.
Therefore, TDCI can correctly model Rabi oscillations.\cite{Peng2018}
The TDCI wave function may be systematically improved to include static electron correlation using an active space expansion or dynamic correlation via typical truncated CI.
As in time-independent electronic structure theory, time-dependent wave function methods based on coupled cluster theory or the algebraic diagrammatic construction provide a size-extensive and efficient description of dynamic correlation in comparison to truncated CI.\cite{sonk_2011,huber_2011,kats_2011,waelz_2012,kvaal_2012,Nascimento2016,Nascimento2017,ruberti_2018,pederson_2019,koulias_2019,cooper_2021,yuwono_2023,ofstad_2023}

Maitra and co-workers resolved the issue that RT-TDDFT has with Rabi oscillations by reformulating it in the response regime.
This method is called response-reformulated TDDFT (RR-TDDFT) and propagates the electronic coefficients, obtained from LR-TDDFT, in a many-electron basis.
In doing this, non-perturbative dynamics, including Rabi oscillations, agree with the exact results.\cite{Dar2024}  
In principle, given the exact, history-dependent functional, this method is exact. 

Another advantage of time-dependent electronic structure methods is that they are great partners for nonadiabtic dynamics simulations, such as Ehrenfest dynamics.\cite{Kosloff1988, Li2005a, Moss2009, slavicek2018_rttdft_ehrenfest, Lacombe2021, Deumal2024, Suchan2025}
Maitra and coworkers further showed that using a special case of RR-TDDFT, where the electronic coefficients obtained from LR-TDDFT are propagated in time, in combination with Ehrenfest dynamics, is superior to the standard RT-TDDFT approach.\cite{Lacombe2021}  As in the case of Rabi oscillations and peak shifting, the failure of RT-TDDFT is related to the non-linearity of the Kohn-Sham operator, and is difficult to avoid without knowledge of the exact, frequency-dependent functional.

With the goal of developing an accurate and efficient time-dependent electronic structure tool for use in simulations of nonadiabatic dynamics, we have implemented the real-time Tamm-Dancoff approximation (RT-TDA).
In RT-TDA, the amplitudes obtained from LR-TDDFT, within both the Tamm–Dancoff and adiabatic approximations, are propagated in real-time.
This is an approach similar to RR-TDDFT\cite{Dar2024}, with two key differences.  First, RR-TDDFT does not invoke the TDA. 
Secondly, RR-TDDFT  couples the excited states through quadratic response\cite{gross1996_1, Casida1995, Tavernelli2009, Parker2017}, whereas RT-TDA couples excitations via the molecular dipole operator, as discussed in more detail below.
These design decisions are made for computational efficiency: eliminating de-excitations via the TDA, avoiding quadratic response, and employing GPU-acceleration.

This paper is structured as follows: In section \ref{sec:methods}, we lay out the details of the RT-TDA theory and the present implementation.
In section \ref{sec:results-discussion}, we present the results of simulations that we obtained using RT-TDA and discuss them.
Finally, in section \ref{sec:conclusions}, we draw conclusions about our method and give an outlook for further advances with RT-TDA.

\section{Theory and Methods}
\label{sec:methods}

In this section, we describe our graphics processing unit (GPU)-accelerated implementation of the real-time Tamm–Dancoff approximation (RT-TDA). 
This method is implemented within TeraChem,\cite{terachem1,terachem2} a GPU-accelerated quantum chemistry software package.
In RT-TDA, the electronic dynamics are represented by a real-time propagation of the amplitudes obtained from LR-TDDFT, within both the Tamm–Dancoff and adiabatic approximations.

\subsection{Real-Time Tamm-Dancoff Approximation}

The central equation in TI-LR-TDDFT is the Casida equation\cite{Casida1995} and the Tamm-Dancoff approximation (TDA) is widely used.  
(In this work, for clarity, we refer to traditional time-independent LR-TDDFT and TDA methods as TI-LR-TDDFT and TI-TDA, respectively.)
The Casida equation within the TDA is given by
\begin{equation} 
\label{eq:casida_tda}
\mathbf{AX} = \omega \mathbf{X},
\end{equation}
where \textbf{X} corresponds to the excitation amplitudes in LR-TDDFT and $\omega$ is the transition frequency for the respective excited state.

For a pair of excitations from orbital $i$ to orbital $a$ and from orbital $b$ to orbital $j$, where $i$ and $j$ are occupied Kohn-Sham molecular orbitals (MOs) and $a$ $b$ are unoccupied ones, one element of the \textbf{A} matrix is defined as 
\begin{equation} 
\label{eq:A_matrix_element2}
    A_{ia, jb} = \delta_{ij} \delta_{ab} (\varepsilon_a - \varepsilon_i) + (ia|jb) + (ia|f_{\text{xc}}|jb).
\end{equation}
The index $ia$ (as well as $jb$) follows TeraChem’s virtuals-run-fastest order for single excitations: for each fixed occupied $i$, we run through all virtuals $a$ before incrementing $i$.
In eq. \ref{eq:A_matrix_element2}, $\varepsilon_a$ and $\varepsilon_i$ correspond to the energy eigenvalues of the orbitals, $(ia|jb)$ are the electron repulsion integrals (ERIs), or two-electron integrals, and $(ia|f_{\text{xc}}|jb)$ is the contribution from the exchange-correlation kernel.
These integrals are defined as
\begin{equation}
(ia|jb)=\iint \frac{\phi_i(\mathbf r_1)\phi_a(\mathbf r_1)\phi_j(\mathbf r_2)\phi_b(\mathbf r_2)}{\lvert \mathbf r_1-\mathbf r_2\rvert} d\mathbf r_1 d\mathbf r_2
\label{eq:eri}
\end{equation}
\begin{equation}
(ia|f_{\text{xc}}|jb)=\iint \phi_i(\mathbf r_1)\phi_a(\mathbf r_1)\frac{\delta^2 E_{\text{xc}}}{\delta \rho(\mathbf r_1)\delta \rho(\mathbf r_2)}\phi_j(\mathbf r_2)\phi_b(\mathbf r_2) d\mathbf r_1 d\mathbf r_2.
\label{eq:fxc}
\end{equation}

If the \textbf{A} matrix was to be treated as an effective Hamiltonian, then one element would be represented as 
\begin{equation} \label{eq:A_matrix_element}
A_{ia,jb} = \langle \Phi_i^a | \hat{A} | \Phi_j^b \rangle,
\end{equation}
where $\Phi_i^a$ is a singly-excited configuration state function (CSF) with an excitation from MO $i$ to MO $a$, analogous for $\Phi_j^b$.

Drawing inspiration from Dar, et al.,\cite{Dar2024} we can think of $\mathbf{A}$ as an effective Hamiltonian.  Given the exact exchange-correlation functional, diagonalization of $\mathbf{A}$ yields the energies of the excited states of the system within the adiabatic and Tamm-Dancoff approximations,

\begin{equation}
\mathbf{A}=\mathbf{UDU^\dagger},
\end{equation}
where $\mathbf{U}$ is a unitary matrix and $\mathbf{D}$ is the diagonal matrix of state energies.
These excited state energies correspond to a set of unknown correlated many-electron wave functions, $\{\tilde \Psi_I\}$.  However, given the similarity to CI singles (CIS), it is reasonable to assume that these wave functions are reasonably well approximated by a CIS-like wave function,
\begin{equation}
\Psi_I \approx \sum_{ia} U_{ia,I} \Phi_i^a,
\end{equation}
where $\{\Phi_i^a\}$ are the singly-excited configuration state functions (CSFs).
Alternatively, we may expand the exact wave function in a basis of correlated many-electron functions, $\{\tilde \Phi_i^a\}$, according to,
\begin{equation}
\Psi_I = \sum_{ia} U_{ia,I} \tilde \Phi_i^a.
\end{equation}
Here $\tilde \Phi_i^a \approx \Phi_i^a$, except that $\tilde \Phi_i^a$ reflects the dynamics electron correlation describe by the exchange correlation functional. 
Within this theoretical framework, $\mathbf{A}$ is the Hamiltonian in the basis, $\{\tilde \Phi_i^a\}$.

We now define our RT-TDA wave function in this correlated basis according to
\begin{equation}
\Psi(t) = \sum_{ia} x_i^a(t) \tilde \Phi_i^a,
\end{equation}
This wave function is propagated by numerically solving the time-dependent Schrödinger equation,
\begin{equation}
\label{eq:ode_rt_tda}
i\dot{\mathbf{X}}(t) = \mathbf{A} \mathbf{X}(t) \equiv \boldsymbol{\sigma}.
\end{equation}
The matrix-vector product between the \textbf{A} matrix and the LR-TDDFT amplitudes, \textbf{X}, is represented by $\boldsymbol{\sigma}$.

In order to describe excitation/deexcitation from/to the ground electronic state, we must add an additional element to $\{\mathbf{X}\}$, $\mathbf{X}_0$, corresponding to the Kohn-Sham ground state.  The exactness of the KS ground state gives us the corresponding elements of $\mathbf{A}$; $A_{0,0}$ is the ground state KS energy, and $A_{0,ai}=A_{ai,0}=0$.  The zero off-diagonals can also be arrived at via the KS analog of Brillouin's theorem.  In what follows, assume this extended formalism if not otherwise indicated.

\subsection{Propagation Algorithm}

The time-dependent excitation amplitudes, $\mathbf{X}(t)$, or expansion coefficients, are decomposed into their real and imaginary components as follows, 
\begin{equation}
\label{eq:x_real_imagiary}
\mathbf{X}(t) = \mathbf{q}(t) + i\mathbf{p}(t).
\end{equation}
In our implementation of RT-TDA, the size of either \textbf{q} or \textbf{p} equals to the number of singly excited CSFs $N_{\text{CSFs}}$ plus one (for the ground state KS configuration), where


\begin{equation}
\label{eq:number_of_csfs}
N_{\text{CSFs}} = n_{\text{occ}} * n_{\text{virt}}.
\end{equation}
Here, $n_{\text{occ}}$ is the number of occupied MOs and $n_{\text{virt}}$ the number of virtual ones.
The size of the \textbf{A} matrix is $(N_{\text{CSFs}}+1)^2$.
Therefore, the size of the \textbf{A} matrix scales with $\mathcal{O}(N^4)$, where $N$ is the size of the system (or number of basis functions).

To start a RT-TDA simulation, we obtain initial conditions.
For this, a ground state Kohn-Sham DFT calculation is carried out, from which the MOs are obtained.
These MOs stay fixed for the rest of the simulation.
In this work, all simulations start from the Kohn-Sham ground state of the field free \textbf{A} matrix, \textbf{A}\textsubscript{0}.
This corresponds to setting $q_0(t=0)$ to one and all the other elements in $\mathbf{q}(t=0)$, as well as all the elements in $\mathbf{p}(t=0)$ to zero.
Of course other initial states, stationary or non-stationary, may also be chosen.

Time-dependent Schrödinger equation can be recast in symplectic form,
\begin{align}
    \dot{\mathbf{q}} &= \mathbf{A} \mathbf{p}, \\
    \dot{\mathbf{p}} &= -\mathbf{A} \mathbf{q}.
\end{align}
To propagate the electronic wave function in time, we use the symplectic split operator (SSO) algorithm.\cite{Blanes2006}
We use this algorithm because of its long-term norm conservation that stems from the symplectic symmetry of the integrator.
Further details of the implementation of SSO for time-dependent electronic structure may be found in ref. \citenum{Peng2018}.

We employ a direct CI approach for propagation.\cite{Roos1972, DavidSherrill1999}
To this end, we take advantage of fast GPU-accelerated routine to calculate $\boldsymbol{\sigma}=\mathbf{A}\mathbf{X}$ developed in the TeraChem software package in the context of TI-LR-TDDFT. 
Details about this GPU accelerated TI-LR-TDDFT implementation can be found in ref. \citenum{Isborn2011}.
The computational cost for calculating an element of \textbf{A} is dominated by ERI contractions and evaluations of the exchange-correlation kernel.
The efficacy stems from calculating the two-electron integrals\cite{McMurchie1978, RezaAhmadi1995} (Coulomb and Exchange terms) over primitive basis functions within the atomic orbital (AO) basis directly on the GPU.
Details about the calculation of the integrals in TeraChem can be found in refs. \citenum{ufimtsev_2008_1, ufimtsev_2009_1}.
The NVIDIA CUBLAS library is used for the linear algebra.
To propagate the wave function for one time step, the routine for calculating $\boldsymbol{\sigma}$ is called twice, once for the real part (\textbf{q}) and once for the imaginary part (\textbf{p}).

The electric field in dipole approximation is incorporated in RT-TDA according to
\begin{equation}
\label{eq:e_field}
    \mathbf{A}(t) \mathbf{X}(t) = \left( \mathbf{A}_0 - \boldsymbol{\mu} \cdot \textbf{d}E(t) \right) \mathbf{X}(t) = \boldsymbol{\sigma} - \boldsymbol{\sigma}_{\text{field}}.
\end{equation}
In this equation, \textbf{A}\textsubscript{0} is the unperturbed \textbf{A} matrix from Casida's equation\cite{Casida1995}, $\boldsymbol{\mu}$ is the dipole matrix in the basis of CSFs, $E(t)$ denotes the scalar-valued time-dependent external field strength, and \textbf{d} is a unit vector indicating the direction of field polarization.
Note that $\mathbf{A}_0$ is time-independent, where as $\mathbf{A}(t)$ on the left-hand side of eq. \ref{eq:e_field} is time-dependent, due to the presence of the electric field.
Here, $\boldsymbol{\sigma}$ corresponds to the unperturbed $\sigma$-formation and $\boldsymbol{\sigma}_{\text{field}}$ is the calculation of the electric field which are represented as
\begin{subequations}
\label{eq:sigmas}
\begin{align}
    \boldsymbol{\sigma} = \mathbf{A}_0 \mathbf{X}(t) \label{eq:unperturbed_sigma} \\
    \boldsymbol{\sigma}_{\text{field}} = \boldsymbol{\mu} \cdot \textbf{d}E(t) \mathbf{X}(t). \label{sigma_field}
\end{align}
\end{subequations}

We have implemented four different pulse types that correspond to different expressions for $E(t)$.
These are given by
\begin{subequations}
\label{eq:e_fields}
\begin{align}
E(t) = E_0 \delta(t) \label{eq:delta_pulse} \\
E(t) = E_0 \sin(\omega t) \label{eq:continuous wave} \\
E(t) = E_0 \cdot \exp\left(-\frac{(t - t_0)^2}{2\sigma^2}\right) \cdot \sin(\omega(t - t_0)) \label{eq:transform_limited} \\
E(t) = E_0 \cdot \exp\left(-\frac{(t - t_0)^2}{2\sigma^2}\right) \cdot \sin\left(\left(\omega + \frac{1}{2}\beta(t - t_0)\right)(t - t_0)\right). \label{eq:chirped_pulse}
\end{align}
\end{subequations}
Eq. \ref{eq:delta_pulse} is the $\delta$-(or kick-) laser pulse, eq. \ref{eq:continuous wave} is the continuous wave, eq. \ref{eq:transform_limited} is the transform limited (TL) and eq. \ref{eq:chirped_pulse} is the chirped pulse, with a time-dependent frequency.
In these equations, $E_0$ is the maximum intensity, $\omega$ is the carrier frequency, $\sigma$ is the pulse width ($\sigma={\text{FWHM}}/{2.35482}$), and $\beta$ is the chirp parameter.
The TL pulse and the chirped pulse are Gaussian shaped laser pulses.
This work contains simulations using the $\delta$-pulse, continuous wave, and TL pulse. 

To ensure efficiency in calculating the dipole matrix $\boldsymbol{\mu}$ in eq. \ref{eq:e_field}, our implementation of RT-TDA obtains the dipole matrix in the MO basis. 
Any matrix in the MO basis contains $\mathcal{O}(N^2)$ elements and is therefore much smaller than a matrix in the CSF basis.
The molecular dipole operator is a one-electron operator and we apply the Slater-Condon rules to it.
By doing this, we can map the elements of the matrix from the MO basis to the CSF basis.
The Slater-Condon rules are summarized in the following:
\begin{subequations}
\label{eq:slater_condon_rules}
\begin{align}
    \langle \Phi_0 | \hat{\mu} | \Phi_0 \rangle &= \sum_{m \in \text{occ}} \langle m | \hat{\mu} | m \rangle \label{eq:sc1}\\
    \langle \Phi_0 | \hat{\mu} | \Phi_i^a \rangle &= \sqrt{2}\langle i | \hat{\mu} | a \rangle \label{eq:sc2}\\
    \langle \Phi_i^a | \hat{\mu} | \Phi_i^a \rangle &= \sum_{m \in \text{occ}} \langle m | \hat{\mu} | m \rangle \notag \\
    & \quad = \langle \Phi_0 | \hat{\mu} | \Phi_0 \rangle - \langle i | \hat{\mu} | i \rangle + \langle a | \hat{\mu} | a \rangle \label{eq:sc3}\\
    \langle \Phi_i^a | \hat{\mu} | \Phi_i^b \rangle &= \langle a | \hat{\mu} | b \rangle \label{eq:sc4}\\
    \langle \Phi_i^a | \hat{\mu} | \Phi_j^a \rangle &= - \langle i | \hat{\mu} | j \rangle \label{eq:sc5}\\
    \langle \Phi_i^a | \hat{\mu} | \Phi_j^b \rangle &= 0. \label{eq:sc6}
\end{align}
\end{subequations}
In these equations, $\Phi_0$ is the ground state Kohn-Sham wave function, and $m$ is a general MO that is occupied in the respective CSF.

In RT-TDA the excited electronic states are coupled through the dipole moment operator that couples the excited CSFs.
In contrast to this, in RR-TDDFT the couplings between the excited states are pre-computed with quadratic-response TDDFT.\cite{Dar2024} 
Furthermore, we take advantage of the fact that the elements in the CSF basis that differ by more than one electron are zero (eq. \ref{eq:sc6}).
Due to this protocol, the number of elements of $\boldsymbol{\mu}$ in eq. \ref{eq:e_field}, that need to be calculated, reduces from $\mathcal{O}(N^4)$ to $\mathcal{O}(N^3)$.
In eq. \ref{eq:e_field}, $\boldsymbol{\sigma}$ is accelerated on the GPU.
Since the scaling of $\boldsymbol{\sigma}_{\text{field}}$ (in the same equation) is lower, its calculation is not performance limiting, and therefore is carried out separately on the CPU.

To obtain an absorption spectrum, we excite the initial state with a $\delta$-pulse.
The $\delta$-pulse is on for the first half of the first time-step in the propagation with the SSO.
Therefore, for a real initial wave function, it only affects the derivative of the imaginary part of the wave function at $t=0$.
Then, we calculate the time-correlation function at each time step.
With $\varepsilon$ being a time immediately after the end of the $\delta$-pulse, the time-correlation function is defined
\begin{equation}
\label{eq:time_correlation_function}
R(t) = \mathbf{X}^\dagger(\varepsilon) \mathbf{X}(\varepsilon+t).
\end{equation}
After the end of the simulation, a Fourier transform of the time-correlation function is carried out according to\cite{Mehmood2024}
\begin{equation}
R(\omega) \propto \omega \mathcal{F}[R(t)].
\label{eq:fourier_transform}
\end{equation}
Here, $R(\omega)$ is the spectrum, $\omega$ is the frequency, and $\mathcal{F}[f(t)]$ is the Fourier-transform of $f(t)$.
Prefactors that do not depend on the frequency are left out, because we compute the spectrum in relative units.
With this, we obtain the absorption spectrum, including excitation energies and oscillator strengths.

Furthermore, at each time step, the populations of the excited states can be calculated.
This is done by projecting the time-dependent wave function onto the stationary states that are obtained from time independent (TI-) TDA calculations,
This is given by
\begin{equation}
\label{eq:populations}
P_i(t) = |\mathbf{X}_i^\dagger\mathbf{X}(t)|^2.
\end{equation}
Here, $\mathbf{X}_i$ is the amplitude vector of state $i$.

\section{Results and Discussion}
\label{sec:results-discussion}

Here we present and discuss the results a series of calculations that highlight the utility of our GPU-accelerated RT-TDA approach.

\subsection{Absorption spectrum}

\begin{figure}
\centering
\includegraphics[width=0.6\linewidth]{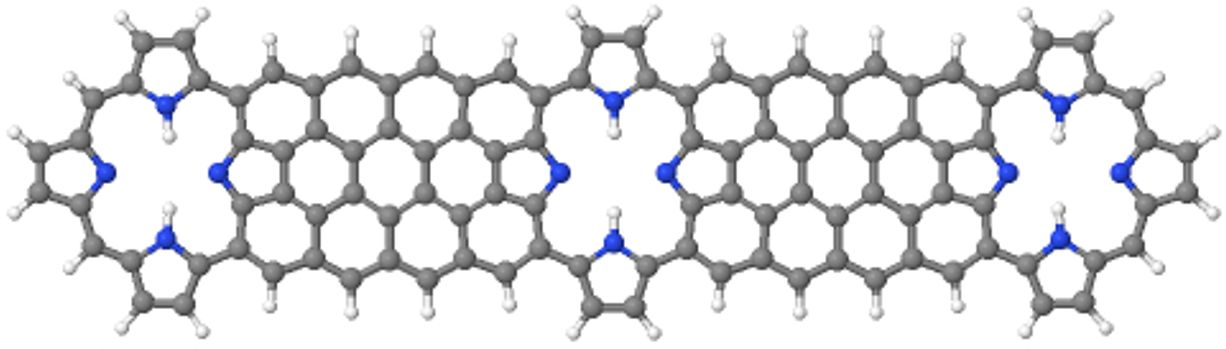}
\caption{The structure of compound 1 (F-Coronene). Carbon, nitrogen, and hydrogen atoms are showin in gray, blue, and white, respectively.}
\label{fig:f_coronene_structure}
\end{figure}

First, we computed the ground-state linear absorption spectrum of the F-Coronene (C\textsubscript{108}H\textsubscript{42}N\textsubscript{12}) molecule with RT-TDA.
Furthermore, we compare it to the results of a TI-TDA reference calculation to assess the accuracy of our method.
The structure of F-Coronene is depicted in Fig. \ref{fig:f_coronene_structure}.
The absorption spectrum was computed as described above using a $\delta$-function laser pulse. 
Three separate simulations were carried out, each for a laser pulse polarized along the $x$, $y$ and $z$ directions of the molecule.
For each simulation, the total time is 3000 a.u. (72.5 fs) and the time step is 0.1 a.u. (2.42 as).
The intensity of the laser pulse is \SI{1e16}{\watt\per\centi\meter\squared}.  Note that the choice of a half-time-step $\delta$-function laser pulse rigorously ensures linear excitation, regardless of the pulse intensity.   The level of theory used is CAM-B3LYP\cite{Lee1988, Becke1993, Yanai2004}/6-31G*.
The geometry was taken from ref. \citenum{Tussupbayev2015}, where it was optimized at the B3LYP\cite{Lee1988, Becke1993}/6-31G* level of theory.
The RT-TDA simulation took one week on three NVIDIA A100 GPUs, where one calculation (for the electric field polarized along one specific direction) is running on one single GPU, independent from the other calculations. 

\begin{figure}
\centering
\includegraphics[width=1.0\linewidth]{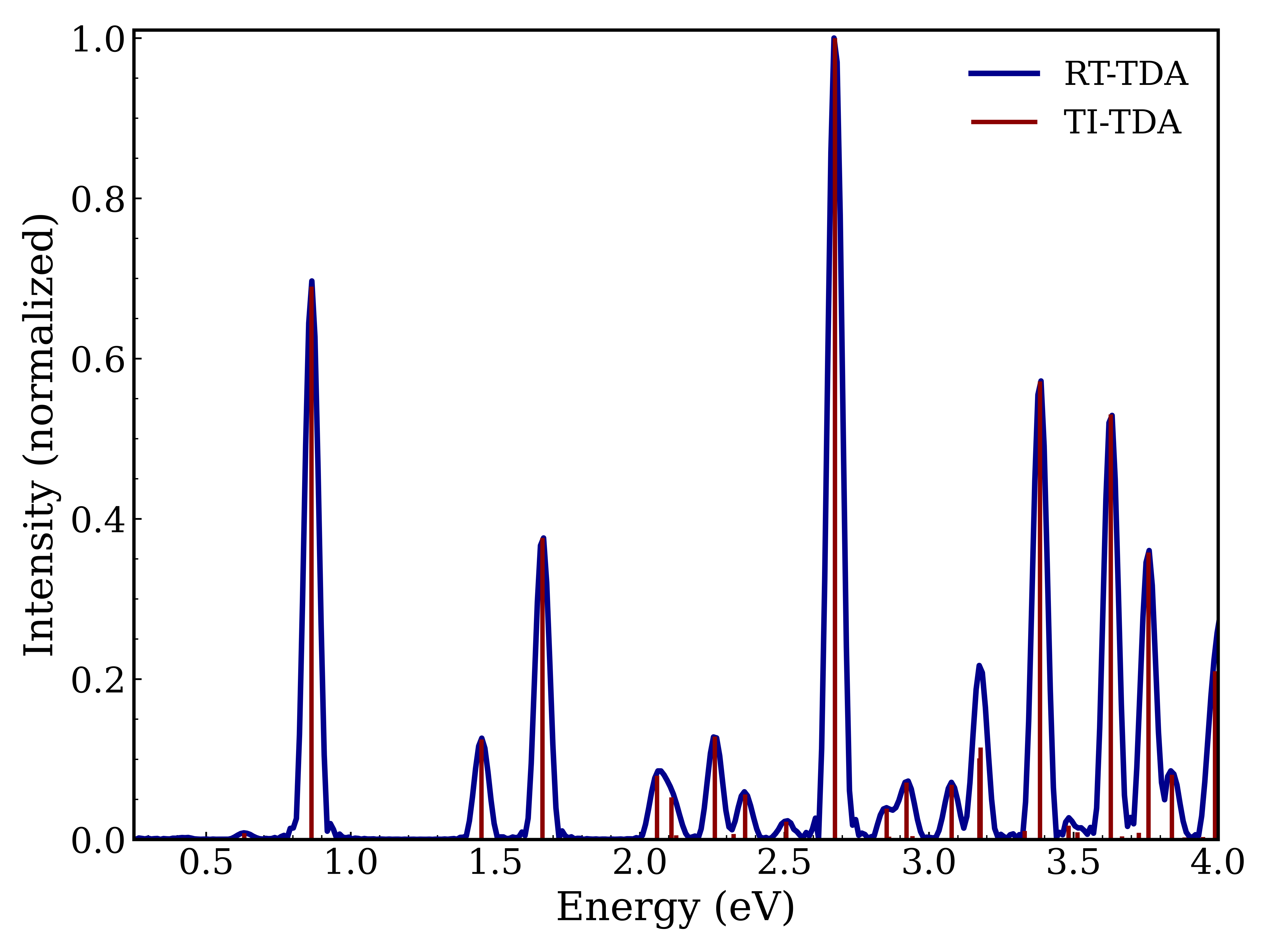}
\caption{Simulated absorption spectrum of the F-Coronene molecule with RT-TDA (blue) compared to reference TI-TDA (red).}
\label{fig:spectrum}
\end{figure}

The spectrum was calculated by taking a fast Fourier transform (FFT) of the time-correlation function (eq. \ref{eq:time_correlation_function}).
The FFT was performed using the Hanning windowing function, mirroring to reduce edge artifacts, and padding of 50,000 zeros.
After the Fourier transform, the angular average of $R(\omega)$ is computed from three simulations with the field polarized in $x$, $y$ and $z$ directions, according to
\begin{equation}
\label{eq:spectrum_average}
    R(\omega) = \frac{1}{3}R_x(\omega) + \frac{1}{3}R_y(\omega) + \frac{1}{3}R_z(\omega).
\end{equation}
Here $R_x(\omega)$ is the component with the laser pulse polarized in the $x$ direction, and so on.
The resulting spectrum, up to 4 eV, which corresponds approximately to the ionization potential (computed with Koopman's theorem), is shown in Figure \ref{fig:spectrum}.
For the above mentioned simulation parameters, the energy resolution is 0.057 eV.
The figure also includes reference calculations of TI-TDA at the same level of theory.
Both the signal obtained from RT-TDA and the sticks obtained from TI-TDA were normalized independently from each other.
Normalization was performed by taking the intensity value of the highest peak and dividing all discrete intensities by this value.
No Gaussian broadening of any of the data was performed.

Table \ref{tab:excitation_energies} compares the excitation energies of the electronic states that correspond to the six peaks with the highest intensity for RT-TDA in \ref{fig:spectrum}.
The table also shows oscillator strengths that were obtained from TI-TDA.
It can be seen that the error in energy for each peak, compared to the reference calculation, is lower than the energy resolution of the chosen simulation parameters (0.057 eV).
This means that we can accurately reproduce excited state energies and absorption spectra obtained with TI-TDA.
The heights of the sticks (obtained from TI-TDA) in Fig. \ref{fig:spectrum} correspond to the calculated oscillator strengths.
The normalized peaks of the RT-TDA spectrum match the heights of the normalized sticks from TI-TDA.
Therefore, we achieve excellent agreement between the two methods and we assert that RT-TDA can correctly reproduce not only excitation energies, but also oscillator strengths from TI-TDA.

\begin{table}[htbp]
\centering
\caption{Excitation energies and oscillator strengths (obtained from TI-TDA) for the six most intense peaks in fig. \ref{fig:spectrum}. Energy values are in eV and oscillator strengths in atomic units. The error in energy is computed as the absolute difference between the two methods.}
\label{tab:excitation_energies}
\begin{tabular}{|c|c|c|c|c|}
\hline
\textbf{Peak \#} & \textbf{RT-TDA (eV)} & \textbf{TI-TDA (eV)} & \textbf{Error (eV)} & \textbf{Osc. (a.u.)} \\
\hline
1 & 0.866 & 0.864 & 0.002 & 3.9143\\
2 & 1.667 & 1.663 & 0.004 & 2.1363\\
3 & 2.672 & 2.675 & 0.003 & 5.6740\\
4 & 3.388 & 3.384 & 0.004 & 3.2434\\
5 & 3.633 & 3.629 & 0.004 & 3.0097\\
6 & 3.762 & 3.759 & 0.003 & 2.0324\\
\hline
\end{tabular}
\end{table}

\subsection{Simulation of two-photon absorption}

In this section, we show that RT-TDA can simulate nonlinear multi-photon processes, enabling access to dark excited states that are out of reach through single-photon excitation. 
This capability arises from the coupling of excited CSFs through the molecular dipole moment operator (eqs. \ref{eq:sc1} -- \ref{eq:sc5}), which allows the system to absorb multiple photons and populate states not directly coupled to the ground state.

\begin{figure}
\centering
\includegraphics[width=0.5\linewidth]{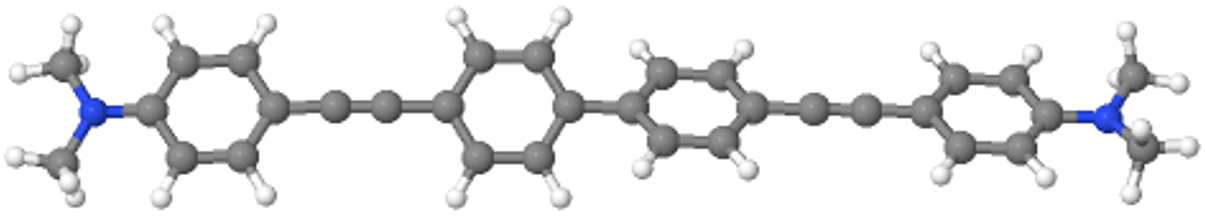}
\caption{The structure of compound 2. Carbon, nitrogen, and hydrogen atoms are show in in gray, blue, and white, respectively.}
\label{fig:dye_structure}
\end{figure}

To investigate the nonlinear two-photon absorption (2PA), we carry out simulations with the dye molecule depicted in Fig. \ref{fig:dye_structure} (compound 2, C\textsubscript{32}H\textsubscript{28}N\textsubscript{2}).
The structure was taken from ref. \citenum{Durden2022}, where it was optimized at the CAM-B3LYP/D95V level of theory.

It was shown in experiment\cite{Pawlicki2009} and with TD-CISNO-CASCI calculations\cite{Durden2022} that this molecule undergoes efficient nonresonant 2PA to the S\textsubscript{2} state.
First, we compute the excitation energies and transition dipole moments (TDMs) of the first two excited states of the molecule with TI-TDA at CAM-B3LYP/D95V level of theory.
The energies are given by 3.953 eV and 4.480 eV, respectively.
The TDM between the ground state and S\textsubscript{1} state is 6.0108 a.u.  The TDM between the S\textsubscript{1} state and S\textsubscript{2} state is similarly large: 5.1108 a.u.  The transition between the ground state and S\textsubscript{2}, however, is essentially dark, with a TDM of 0.0045 a.u. 
Population transfer into S\textsubscript{2} may, in principle, proceed through various pathways via nonlinear multiphoton absorption, however. 
Based on the computed transition dipole moments, one likely pathway in this case is through the bright S\textsubscript{1} state, which is strongly coupled to both the ground state and S\textsubscript{2}.

\begin{figure}
\centering
\includegraphics[width=1.0\linewidth]{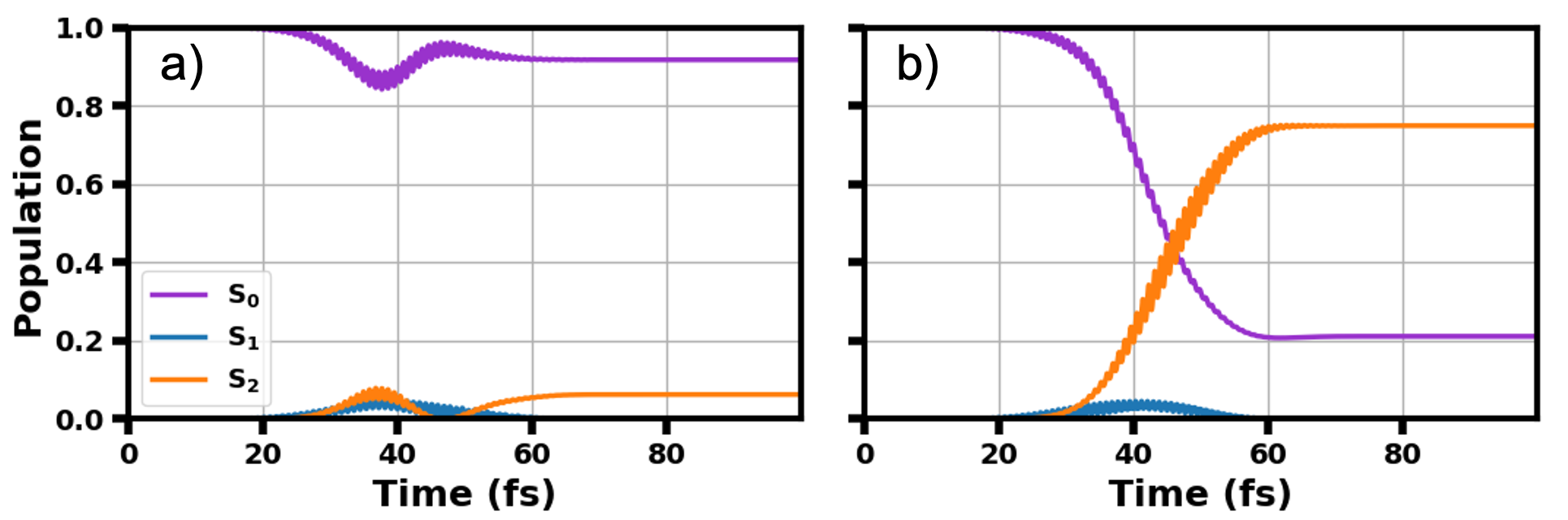}
\caption{Simulation of nonresonant 2-photon absorption of compound 2, with laser frequency tuned to roughly half the excitation energy of the S\textsubscript{2} state (excitation energy = 4.480 eV). The FWHM of the laser pulse is 30 fs. The intensity of the laser pulses is \SI{5e-11}{\watt\per\centi\metre\squared}, and the
frequency is tuned to either a) exactly half of the excitation energy (2.240 eV; \SI{5.416e14}{\hertz}), or b) 0.080 eV higher in energy (2.320 eV; \SI{5.610e14}{\hertz}). Populations are obtained by projecting the time-dependent wave function onto the states that are obtained from TI-TDA at the same level of theory (see eq. \ref{eq:populations}).}
\label{fig:2pa_dye_nonres}
\end{figure}

In the next step, we carry out a RT-TDA simulation (at CAM-B3LYP/D95V level of theory) with a TL laser pulse.
The frequency of the laser pulse was set to half the excitation energy of the S\textsubscript{2} state (2.240 eV, or \SI{5.416e14}{\hertz}).
The maximum intensity of the laser pulse is \SI{5e-11}{\watt\per\centi\metre\squared} and the FWHM of the pulse is 30 fs.
Fig. \ref{fig:2pa_dye_nonres}a shows the population dynamics of this RT-TDA simulation.
It can be seen that the population of the target 
S\textsubscript{2} state remains below approximately 5\% throughout the simulation, indicating that only a small fraction of the total population is transferred.
The S\textsubscript{0} population stays above 90\% during the entire simulation, which means that almost the entire wave function remains in the ground state.
The low maximum population of S\textsubscript{2} demonstrates that the transfer is inefficient. 
Moreover, at around $t \approx 38\ \mathrm{fs}$, the population in S\textsubscript{2} starts to decrease while the S\textsubscript{0} population rises, showing that the system is relaxing back toward the ground state. 
Though we have carefully tuned the laser frequency so that the S$_0$ to S$_2$ transition is two-photon resonant, the inefficient, oscillatory dynamics we observe are consistent with the laser field being off-resonance.

Fig. \ref{fig:2pa_dye_nonres}b presents population dynamics for a simulations where the carrier frequency of the laser pulse was increased by 80 meV, to 2.320 eV, or \SI{5.610e14}{\hertz}, compared to the frequency in Fig. \ref{fig:2pa_dye_nonres}a.
Otherwise the same pulse shape, simulation parameters, and level of theory were used.
Here, it can be seen that the S\textsubscript{2} population reaches its maximum of approximately 75\% at $t \approx 60\ \mathrm{fs}$.
Furthermore, the population stays there until the end of the simulation.
Therefore, we assert that an efficient population transfer (as seen in fig. \ref{fig:2pa_dye_nonres} b)) requires a shift in laser frequency for this case.
The detuning of the transition appears to be due to the Autler–Townes effect (or AC Stark effect)\cite{Autler1955, Schuda1974, TralleroHerrero2005, Clow2008}, the analog of the Stark effect for strong oscillatory fields.
A future work will do a comparative study of the ability of different methods to describe the Autler-Townes effect.
The results presented in this section demonstrate that RT-TDA can accurately capture nonlinear excitation pathways, including multiphoton absorption and dynamic detuning effects.

\subsection{Rabi Oscillations}

As mentioned above, it is well established that RT-TDDFT has limitations in terms of describing Rabi oscillations accurately. 
These shortcomings arise on the one hand from the detuning of excitation energies, as the Kohn–Sham matrix evolves over time\cite{fuks_2011,Provorse2015}; and on the other hand because the method's mean-field character leads to artificial two-electron transitions between orbitals, resulting in unphysical two-electron Rabi oscillations.\cite{Habenicht2014}
In contrast, wave function methods such as TD-CASCI, which are based on a linear, many-electron Hamiltonian, can correctly model Rabi Oscillations.\cite{Peng2018}
More recently, Maitra and coworkers have shown that RR-TDDFT can also correctly model Rabi oscillations, which may again be attributed to the linearity of the propagation.\cite{Dar2024}

\begin{figure}
\centering
\includegraphics[width=1.0\linewidth]{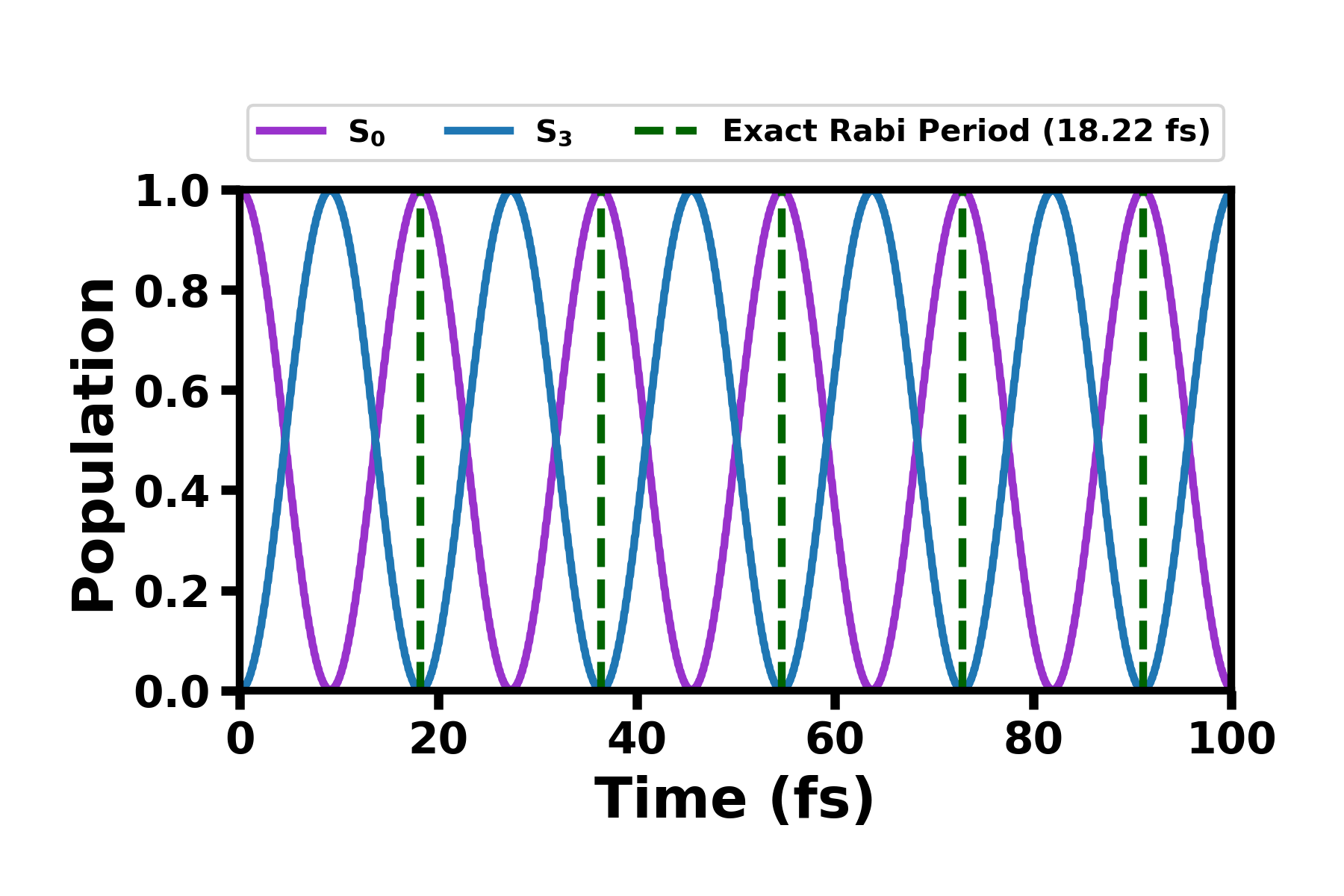}
\caption{Populations of the ground state and S\textsubscript{3} state ($\pi \rightarrow \pi^*$ excitation character) as a function of time. Populations are obtained by projecting the time-dependent wave function onto the states that are obtained from TI-TDA at the same level of theory (see eq. \ref{eq:populations}).}
\label{fig:rabi}
\end{figure}

Since Rabi oscillations are an exactly solvable problem, we demonstrate the accuracy of RT-TDA by reproducing them.
To this end, we simulate the response of an ethylene molecule to a CW wave field, polarized along the C\!=\!C bond.
The geometry for the ethylene molecule was taken from ref. \citenum{Peng2018}, where it was optimized at the B3LYP/6-31G** level of theory.
The level of theory for the RT-TDA simulation is CAM-B3LYP/STO-3G and the time step is 0.05 a.u. (1.2 as).
The maximum intensity of the field  is \SI{1e12}{\watt\per\centi\metre\squared} and the applied frequency is \SI{2.848e15}{\hertz} (11.78 eV). 
The frequency corresponds to the excitation energy of the S\textsubscript{3} state ($\pi \rightarrow \pi^*$ transition), computed with TI-TDA at the same level of theory.
This state is the lowest bright excited state and has a transition dipole moment of 1.5624 a.u..
The exact Rabi period is calculated using the expression \(T = 2\pi\hbar / (\mu E_0)\) which is derived from the well-known Rabi frequency\cite{Tannor2006} \(\Omega = \mu E_0 / \hbar\), where $\mu$ is the transition dipole moment between the two states (in the direction at which the electric field is applied) and $E_0$ is the electric field strength.
The results of our RT-TDA simulation are depicted in Figure \ref{fig:rabi}. Table \ref{tab:rabi_periods} shows the Rabi periods for the first five cycles of our simulation, compared to the exact result.
It can be seen that we achieve excellent agreement.
Furthermore, there is a complete population inversion at each cycle. 
Therefore, we show that RT-TDA does not suffer from the artifactual dynamical detuning present in RT-TDDFT.

\begin{table}[htbp]
\centering
\caption{Comparison of Rabi oscillation periods computed with RT-TDA and the exact solution. The error in the period is computed as the absolute difference between RT-TDA and the exact result.}
\label{tab:rabi_periods}
\begin{tabular}{|c|c|c|c|}
\hline
\textbf{Oscillation} & \textbf{RT-TDA (fs)} & \textbf{Exact (fs)} & \textbf{Error (fs)} \\
\hline
First  & 18.26 & 18.22 & 0.04 \\
Second & 36.51 & 36.44 & 0.07 \\
Third  & 54.59 & 54.66 & 0.07 \\
Fourth & 72.85 & 72.88 & 0.03 \\
Fifth  & 91.10 & 91.10 & 0.00 \\
\hline
\end{tabular}
\end{table}

\subsection{Convergence of the SSO Integrator}

\begin{figure}
\centering
\includegraphics[width=1.0\linewidth]{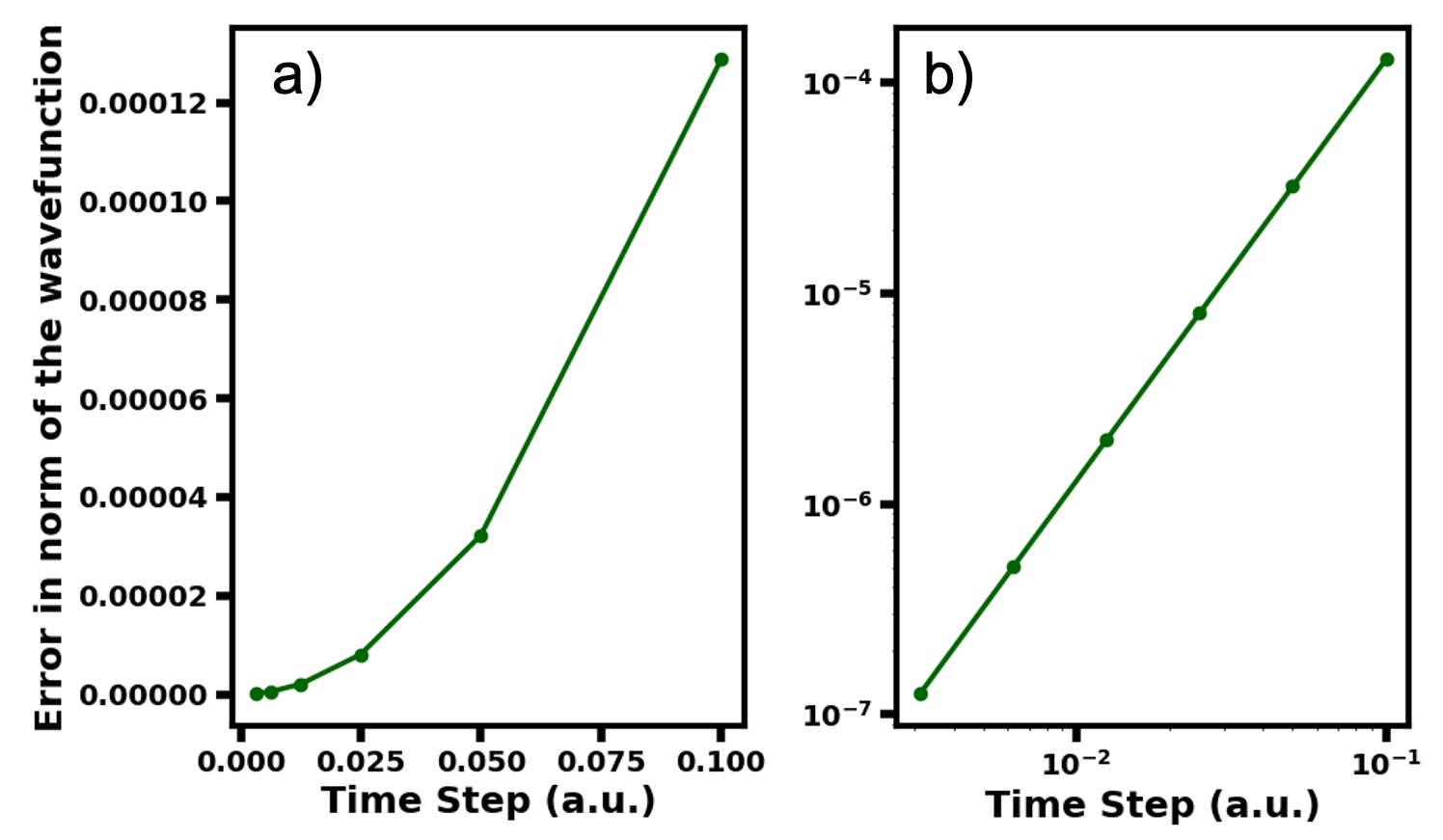}
\caption{Convergence of the SSO integrator. The same data is plotted in a a) linear plot and b) log-log plot.}\
\label{fig:sso_convergence}
\end{figure}

We carried out a convergence study to verify the numerical accuracy and stability of the second-order SSO integrator\cite{Blanes2006} used in RT-TDA.
For this, we ran multiple simulations of the CO molecule at CAM-B3LYP/STO-3G level of theory.
The C--O bond distance was set to the experimental value of 1.128 Å\cite{Huber1979}.
In these simulations, we initialize the real part of the \textbf{X} vector with the converged amplitudes of the first electronically excited state, obtained with TI-TDA at the same level of theory. 
We let the simulation run with a fixed simulation time of 1.2 fs (50 a.u.), varying the time step.
Short-term fluctuations in the norm are expected, but the goal is to achieve long-term conservation of the norm.
Due to this, we use the average of the error in norm throughout the whole simulation as a metric.
The results of these simulations can be seen in figure \ref{fig:sso_convergence}.
Both graphs show the same data, but fig. \ref{fig:sso_convergence}a is a linear plot while fig. \ref{fig:sso_convergence}b is a log-log plot.
It can be seen that the error decreases with decreasing step size, which confirms a successful integration in time.
Furthermore, in fig. \ref{fig:sso_convergence} b), the slope of the graph is two.
This verifies the second-order accuracy of the SSO for the time integration of the ODE in RT-TDA.

\subsection{Timings}

To assess the efficiency of our GPU accelerated implementation, we carried out benchmark calculations of three different sized molecules.
These molecules are C\textsubscript{16}H\textsubscript{10}N\textsubscript{2}O\textsubscript{2} (indigo dye), C\textsubscript{32}H\textsubscript{28}N\textsubscript{2} (compound 2, in fig. \ref{fig:dye_structure}), and C\textsubscript{108}H\textsubscript{42}N\textsubscript{12} (F-Coronene, compound 1, in fig. \ref{fig:f_coronene_structure}).
For better comparison, all the simulations in this section were carried out with the same parameters of a prototypical TL laser pulse.
The total simulation time is 2.4 fs (100 a.u.) and the time step is 1.2 as (0.05 a.u.).
Therefore, each simulation has 2000 steps.
In one step, the formation of $\mathbf{\sigma}$ and the electric field ($\boldsymbol{\sigma}_{\text{field}}$) occur twice, once for the real part of the \textbf{X} vector and once for its imaginary part.
For each molecule, the pulse is polarized along the long axis of the respective molecule (such that it aligns with the direction of the transition dipole moment between the ground state and the lowest brightest excited state).
The pulses have a frequency of \SI{6.045e14}{\hertz} (2.500 eV), a maximum intensity of \SI{5.0e11}{\watt\per\centi\meter\squared}, and a FWHM of 0.91 fs.
The timings for these three simulations on two different GPUs are summarized in Tab. \ref{tab:timings}.
Each value in this table represents the total wall-clock time spent on a specific computational task ($\boldsymbol{\sigma}$-vector formation or electric field formation) over the entire simulation. 
For a 2.4 fs simulation with 2000 time steps, each of these tasks is performed twice per step (once for the real part of the vector and once for the imaginary part), leading to 4000 recorded timings. 
Each value reported in the table is the sum of these 4000 timings.
In all cases, $\boldsymbol{\sigma}$ formation on the GPU requires more than four times longer than formation of the field on the CPU.
This is because the scaling of $\boldsymbol{\sigma}$ is higher than the field formation.
Therefore, calculating the field on the CPU is not performance limiting for systems up to a size of a few thousand basis functions.

Furthermore, it can be seen that carrying out the calculation on a better hardware, such as the NVIDIA A100 GPU, accelerates the time integration significantly.
This is a promising result since the development of newer and more efficient GPU hardware systems is ongoing and this will make it feasible to study even larger systems with RT-TDA.

\begin{table}[htbp]
\centering
\caption{Breakdown of the total wall-clock times (in seconds) required for specific computational components of a 2.4 fs RT-TDA simulation (2000 time steps) on different GPUs and for different molecular systems. For each system, we report the cumulative time spent on $\boldsymbol{\sigma}$-vector formation and on electric field formation separately. Each entry is the total time for that task over all 2000 time steps, including both real and imaginary components (4000 operations in total for each task). HA stands for heavy atom (carbon, nitrogen and oxygen) and BF stands for basis function.}
\label{tab:timings}
\resizebox{\textwidth}{!}{
\begin{tabular}{|c|c|c|c|c|c|c|}
\hline
\textbf{GPU} & \multicolumn{2}{c|}{\textbf{C\textsubscript{16}H\textsubscript{10}N\textsubscript{2}O\textsubscript{2} (Indigo)},} & \multicolumn{2}{c|}{\textbf{C\textsubscript{32}H\textsubscript{28}N\textsubscript{2}},} & \multicolumn{2}{c|}{\textbf{C\textsubscript{108}H\textsubscript{42}N\textsubscript{12} (F-Coronene)},} \\
& \multicolumn{2}{c|}{20 HAs, 136 e\textsuperscript{-}, 320 BFs} & \multicolumn{2}{c|}{34 HAs, 234 e\textsuperscript{-}, 566 BFs} & \multicolumn{2}{c|}{120 HAs, 774 e\textsuperscript{-}, 1884 BFs} \\
\cline{2-7}
& \textbf{$\boldsymbol{\sigma}$ (GPU)} & \textbf{field (CPU)} & \textbf{$\boldsymbol{\sigma}$ (GPU)} & \textbf{field (CPU)} & \textbf{$\boldsymbol{\sigma}$ (GPU)} & \textbf{field (CPU)} \\
\hline
\textbf{A100} & 1094 & 33.14 & 2173 & 180.6 & 32650 & 6984 \\
\hline
\textbf{V100} & 1328 & 26.70 & 2927 & 171.0 & 49540 & 5868 \\
\hline
\end{tabular}
}
\end{table}

\section{Conclusions}
\label{sec:conclusions}

In this work, we have realized and presented the GPU accelerated implementation of RT-TDA.
This is an efficient and accurate tool to simulate electron dynamics in complex molecules.
RT-TDA is based on the real-time propagation of the LR-TDDFT amplitudes within the TDA and adiabatic approximation.
We showed that our approach is accurate on the one hand by correctly modeling Rabi Oscillations.
On the other hand, we compared the RT-TDA ground state absorption spectrum to the excitation energies and oscillator strengths from TI-TDA, where we achieve excellent agreement.
Therefore, RT-TDA provides an efficient and robust way of computing absorption spectra, where convergence issues of stationary states can be avoided through a time propagation.
The computation of the RT-TDA absorption spectrum of the F-Coronene molecule from a 72.5 fs simulation took 1 week on a single NVIDIA A100 GPU (three calculations running naively parallel).
Furthermore, we showed that RT-TDA can access all the states that are present in LR-TDDFT within the adiabatic approximation (singly excited states) via nonlinear multiphoton absorption.
In addition to this, RT-TDA is found to describe the AC Stark effect in these simulations.
In the future, we plan to use RT-TDA as the electronic structure component of mixed quantum–classical nuclear dynamics simulations.
Specifically, we plan on coupling RT-TDA to the decoherence-corrected Ehrenfest dynamics code TAB-DMS (to-a-block dense-manifold-of-states)\cite{Esch2020, Esch2020a, Esch2021, Liang2024}, where it will contribute as an efficient way of treating dynamic electron correlation.

\section*{Acknowledgments}
We gratefully acknowledge Jiří Suchan and Arshad Mehmood for fruitful discussion.
This work was supported
by grant award DE-SC0021643
from the Solar Photochemistry program
of the Office of Basic Energy Sciences,
U.S. Department of Energy.
We also gratefully acknowledge funding
from the Institute
for Advanced Computational Science
at Stony Brook University.
This work used Expanse GPU at the San Diego Supercomputer Center (SDSC) through allocation CHE140101 from the Advanced Cyberinfrastructure Coordination Ecosystem: Services and Support (ACCESS) program, which is supported by National Science Foundation grants \#2138259, \#2138286, \#2138307, \#2137603, and \#2138296.

\noindent \textbf{Competing Interests:} The authors declare that they have no competing interests.

\noindent \textbf{Data Availability:} The data are available from the Corresponding Author upon request.

\section*{Supporting Information Available}
The optimized geometries used in this work are presented in the Supporting Information document.


\newpage

\bibliography{bibliography}

\end{document}